\documentclass[10pt]{iopart}

\usepackage{graphicx}
\begin{document}

\title[High-sensitivity Kinetic Inductance Detectors for CALDER]{High-sensitivity Kinetic Inductance Detectors for CALDER}

\author{A. D'Addabbo $^1$\footnote{Author's email: antonio.daddabbo@lngs.infn.it}, F. Bellini $^{2,3}$, L. Cardani $^{3}$, N. Casali $^{3}$, M. G. Castellano $^4$, I. Colantoni $^4$, C. Cosmelli $^{2,3}$, A. Cruciani $^{3}$, S. Di Domizio $^{5,6}$, M. Martinez $^{2,3}$, C. Tomei $^{3}$, M. Vignati $^{3}$}

\address{$^1$ INFN - Laboratori Nazionali del Gran Sasso (LNGS), Via Giovanni Acitelli 22, 67010, Assergi (AQ) - Italy
\newline $^2$ Dipartimento di Fisica - Sapienza Università di Roma, Piazzale Aldo Moro 2, 00185, Roma - Italy
\newline $^3$ INFN - Sezione di Roma, Piazzale Aldo Moro 2, 00185, Roma - Italy
%\newline $^4$ Physics Department - Princeton University, Washington Road, 08544, Princeton - NJ, USA
\newline $^4$ Istituto di Fotonica e Nanotecnologie - CNR, Via Cineto Romano 42, 00156, Roma - Italy
\newline $^5$ Dipartimento di Fisica - Università degli Studi di Genova, Via Dodecaneso 33, 16146, Genova - Italy
\newline $^6$ INFN - Sezione di Genova, Via Dodecaneso 33, 16146, Genova - Italy}
%\ead{submissions@iop.org}
\vspace{10pt}
\begin{indented}
\item[]March 2017
\end{indented}

\begin{abstract}
Providing a background discrimination tool is crucial for enhancing the sensitivity of next-generation experiments searching for neutrinoless double-beta decay. The development of high-sensitivity ($<$ 20 eV RMS) cryogenic light detectors allows simultaneous read-out of the light and heat signals and enables background suppression through particle identification. 
The Cryogenic wide-Area Light Detector with Excellent Resolution (CALDER) R\&D already proved the potential of this technique using the phonon-mediated Kinetic Inductance Detectors (KIDs) approach. The first array prototype with 4 Aluminum KIDs on a 2 $\times$ 2 cm$^2$ Silicon substrate showed a baseline resolution of 154 $\pm$ 7 eV RMS. Improving the design and the readout of the resonator, the next CALDER prototype featured an energy resolution of 82 $\pm$ 4 eV, by sampling the same substrate with a single Aluminum KID. %Preliminary results show that further improvements in resolution can be reached using materials with lower T$_c$ and higher kinetic inductance content. 
\end{abstract}

%Large-mass arrays of bolometers proved to be good detectors for neutrinoless double beta decay (0?DBD) and dark matter searches. CUORE and LUCIFER are bolometric 0?DBD experiments that will start to take data in 2015 at Laboratori Nazionali del Gran Sasso in Italy. The sensitivity of CUORE could be increased by removing the background due to ? particles, by detecting the small amount of C?erenkov light (100eV) emitted by the ?s? signal and not by ?s. LUCIFER could be extended to detect also Dark Matter, provided that the background from ? /? particles (100 eV of scintillation light) is discriminated from nuclear recoils of about 10 keV energy (no light). We have recently started a new project to develop light detectors for CUORE, LUCIFER and similar bolometric experiments. The aim is to develop detectors with an active area of 5x5 cm2 (the face of bolometric crystals), operating at 10 mK, and with an energy resolution at the baseline below 20 eV RMS. We have chosen to develop phonon-mediated detectors with KID sensors. We are currently developing the first prototypes that we plan to test and calibrate during the summer.

%\maketitle
\ioptwocol

\section{Introduction}

%The suppression of the background is crucial for bolometric experiments searching for neutrinoless double-beta decay. 
The next-generation ton-scale experiments searching for neutrinoless double beta decay must be sensitive to a Majorana neutrino mass as low as 10 meV. %The development of high-sensitivity cryogenic light detectors would enable background suppression exploiting the different light yield of different particles.The development of high-sensitivity cryogenic detectors for optical-UV photons is of primary interest for next-generation experiments searching for rare interactions in Neutrino and Dark Matter Physics. 
Among them, CUORE (Cryogenic Underground Observatory for Rare Events) \cite{CUORE} is an array of 988 TeO$_2$ bolometers %searching for neutrinoless double beta decay in $^{130}$Te
currently in its detector commissioning phase at Laboratori Nazionali del Gran Sasso, in Italy. It features a five-years projected sensitivity of 50-130 meV at 90\% C.L. The removal of the background from $\alpha$ radioactivity would push the experiment sensitivity in the Majorana neutrino mass signal region. This is possible by detecting the tiny amount of Cherenkov light emitted only by the $\beta$ signal in coincidence with the heat release in a bolometer \cite{CasaliEPJC}.%The amount of light detected is so far limited to only 100 eV, requiring low-noise cryogenic light detectors.
 To match the requirements of next-generation experiments, the light detectors must feature energy resolution better than 20 eV RMS, large active area (tens of cm$^2$), and the possibility of operating about 1000 channels in a wide temperature range (5 to 20 mK) \cite{Artusa}\cite{CUPID}\cite{CUPID2}.

Among the proposed technologies, Kinetic Inductance Detectors (KIDs) stand out for their natural multiplexed read-out and excellent intrinsic energy resolution \cite{KID_Nature}. To overcome their poor active surface ($\sim$ mm$^2$) KIDs can be coupled to a much wider insulating substrate: photons interacting in the substrate are converted into phonons and finally absorbed by KIDs \cite{Swenson}\cite{Moore}.

The CALDER project \cite{CALDER2015} aims at developing a small prototype experiment consisting of TeO$_2$ bolometers coupled to new light detectors based on KIDs. The R\&D is focused on the light detectors that could be implemented in a next-generation neutrinoless double-beta decay experiment.

\section{Methods}

The prototypes of light detectors developed by the CALDER collaboration consist in 300 to 380 $\mu$m thick, 2 $\times$ 2 cm$^2$ wide high-resistivity \textit{Si}(100) substrates sampled by one to four KID sensors, with a LEKID design \cite{Doyle}. The superconducting resonators are made with 40nm-thick or 60nm-thick Aluminum, patterned by electron beam lithography on a single film deposited using electrongun evaporator \cite{Colantoni}. The detectors are design to feature high quality factors and to resonate at about 2.5 GHz, as expected by SONNET simulations.%For the single pixel design, the detector is optimized to maximize the active area (4.0 mm$^2$, total filling ratio of 2.3\%). 

The chip is mounted on a proper holder through four PTFE (Polytetrafluoroethylene) supports with a contact area of few mm$^2$ to reduce thermal contact and so phonon escape. The detectors are cooled down by a wet $^3$He/$^4$He dilution refrigerator with base temperature of 10 mK.
The readout electronics - NIKEL - has been developed by LPSC of Grenoble within the NIKA collaboration. It is and FPGA-based board that allows to read out up to eight pixels with a data rate of 0.5-2MHz \cite{Bourrion2011}\cite{Bourrion2013}. For a typical bias power P$_{in}\sim$-70 dBm, overall performances of the readout chain reach -120 dBc/Hz.
The detectors can be illuminated by a multimode optical fiber, coupled to a fast room-temperature LED ($\lambda$ = 400nm). Calibration with X-rays are performed facing $^{55}$Fe (5.9 keV) and $^{57}$Co (6.4 keV, 14.4 keV) radioactive sources. 

The analysis pipeline \cite{Casali} we developed allows us to evaluate the main resonators parameters (quality factors, asymmetry), using a unique fit that corrects all the main distortions due to the read-out chain and the bias power. Energy calibration of the devices is performed evaluating broken Cooper pairs as an equivalent of a thermal induced variation \cite{Cardani2015}. The degradation of the resolution due to the noise is minimized using an optimum filter technique.

\section{Results}

The first CALDER prototype \cite{Cardani2015} obtained by depositing four 40nm thick Al KIDs on a 2 $\times$ 2 cm$^2$, 300 $\mu$m thick Si substrate, reached a combined baseline resolution of 154 $\pm$ 7 eV, and an overall efficiency of 18 $\pm$ 2 \%, remarkable for an indirect detection. In this design, the active area of the resonator was $2.4$ mm$^2$, corresponding to a single KID absorption efficiency of 3.1\% -- 6.1\%, depending on the position of the source.

We characterized the 4-pixel array with coupling quality factor (Q$_c$) values in the range of 6--35 $\times$ 10$^3$. The typical pulse shape shows rise time of 10-30 $\mu$s, which is determined by the athermal phonon propagation into the substrate, and decay time of 200-1000 $\mu$s, that is dominated by quasi-particle recombination time ($\tau_{qp}$). A typical value at the optimum bias power, i.e. the power where the SNR (Signal-to-Noise Ratio) of an optical signal is maximized, is $\tau_{qp} \sim$ 220 $\mu$s. 

\begin{figure}[h]
\centering
\includegraphics[width=7cm,clip]{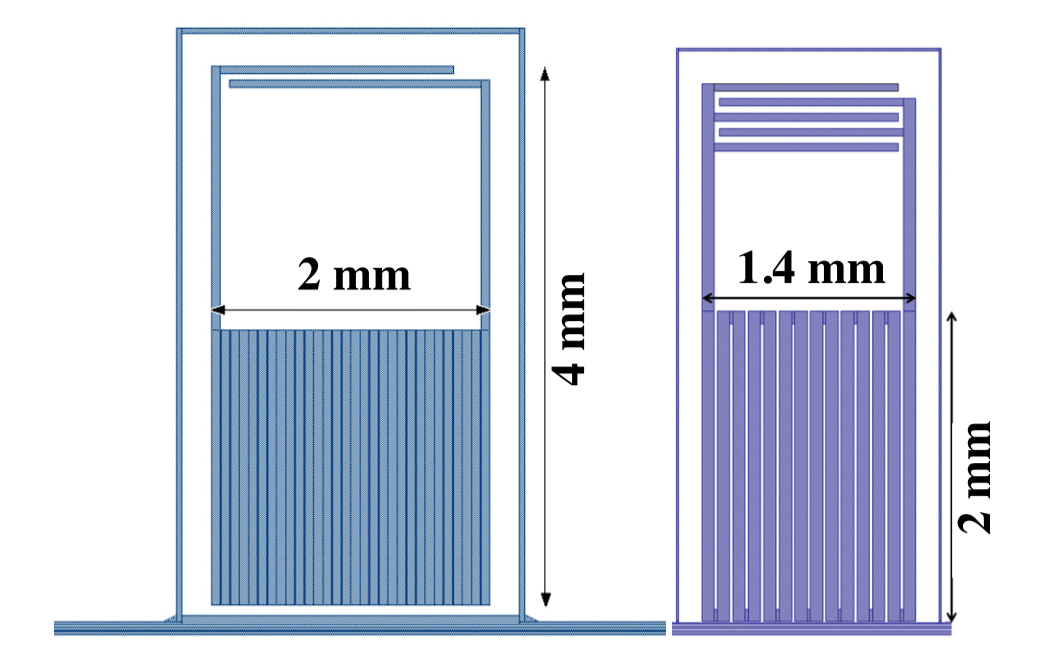}
\caption{Left: diagram of the single-pixel design. The inductor (30 strips of 62.5 $\mu$m $\times$ 2 mm) features an active area of 3.75 mm$^2$ (4.0 mm$^2$ including the active region that connects the inductor to the capacitor). To contain the geometrical inductance, we used a gap of 5 $\mu$m between the meanders and we closed the circuit using a capacitor made by only 2 fingers.
Right: diagram of the 4-pixels design. The active area of the inductive meander (2.4 mm$^2$) consists of 14 connected strips of 80 $\mu$m $\times$2mm, with a gap of 20 $\mu$m.
\newline Reproduced with permission from Cardani et al., Appl. Phys. Lett. 110, 033504 (2017). Copyright 2017 AIP Publishing.}
\label{fig1}
\end{figure}

To improve the detector resolution, we tested a single Al KID design in order to avoid cross-talk or competition in absorbing phonons. The single KID was deposited on 2 $\times$ 2 cm$^2$, 380 $\mu$m thick, high resistivity ($>$10 k$\Omega$ $\cdot$ cm) \textit{Si}(100) substrates.  
Then, in order to increase the signal, we first raised the quality factor Q of the resonator ($\frac{1}{Q} = \frac{1}{Q_c} + \frac{1}{Q_i}$): Q$_c$ was raised up to 150  $\times$ 10$^3$ by design, and we used a 60 nm thick film to ensure a high internal quality factor Q$_i$ ($>$ 10$^6$). Then, we enlarged the active area of the KID from to 4.0 mm$^2$, in order to increase the fraction of phonons that can be collected. A comparison of the improved design with the one described in \cite{Cardani2015} is shown in figure~\ref{fig1}.
Finally, we developed a combined amplitude ($\delta$A) and phase ($\delta\phi$) analysis technique  that allowed us to combine the $\delta$A and $\delta\phi$ signals and maximize the SNR \cite{Cardani2017}.
Thanks to these improvements, the single-pixel design detector featured a KID efficiency up to 7.4\% -- 9.4\%, and a noise baseline resolution of $\sigma_E$ = (82 $\pm$ 4) eV, a factor 2 better than the previous 4-pixels prototype described in \cite{Cardani2015}.

\section{Discussion}

%As discussed later in the text, all the tested prototypes featured an excess low-frequency noise consistent with what observed in our first prototype 20
%Therefore, to improve the signal-to-ratio (SNR), we tried to increase the signal. 
Thanks to the high Q resonator, improved geometry and combined readout, we obtained a signal height about a factor 6 larger with respect to the ones obtained with previous prototype, both in $\delta$A and $\delta\phi$. However, all the tested prototypes featured an excess low-frequency noise that can just partially be ascribed to Two Level System. We are currently investigating its origin. On the other hand, the amplitude noise is consistent with the amplifier temperature and much lower than the phase one. For this reason, even if the amplitude signals are $\sim$10 times smaller than the phase ones, the SNRs are similar and both actually concur to enhance the energy resolution.

\section{Conclusions and perspectives}

An improvement of detectors' sensitivity by at least a factor 2.5 is still needed to reach  the goal of the project. Looking at the resolution expected if the noise was dominated by the cold amplifier, $\sigma_{amp}$, as described in \cite{CALDER2015}, further improvements in sensitivity can be obtained by using of lower critical temperature (T$_c$) superconductors and/or higher kinetic inductance content materials. We are currently studying several new superconducting material for our application \cite{Colantoni} (non stoichiometric TiN \cite{Ohya}, Ti-Al bi-layer \cite{Catalano} and Ti-TiN multilayer \cite{Vissers}) that could take advantage of both these factors. %Preliminary tests shown performances remarkably good, pushing the noise resolution below 50 eV.

By the end of 2017, we foresee to conclude the development to bring the energy resolution of the light detectors below 20 eV RMS. In early 2018, we expect to construct a demonstrator, made of a bolometric TeO$_2$ array at Laboratori Nazionali del Gran Sasso, instrumented with CALDERs light detectors.


\section*{References}

\begin{thebibliography}{}

\bibitem{CUORE} A. D'Addabbo et al., EPJ. Web Conf., In Press, arXiv:1612.04276 (2016) %The CUORE and CUORE-0 experiments at LNGS,
\bibitem{CasaliEPJC} N. Casali et al., EPJ C 75, 12 (2015)
\bibitem{Artusa} D.R. Artusa et al., EPJ C 74, 3096 (2014) 
\bibitem{CUPID} The CUPID Interest Group, arXiv:1504.03599 (2015) %CUPID: CUORE Upgrade with Particle IDentification, 
\bibitem{CUPID2} The CUPID Interest Group, arXiv:1504.03612 (2015) %R\&D towards CUPID
\bibitem{KID_Nature} P. K. Day et al., Nature 425, 817 (2003)
\bibitem{Swenson} L. J. Swenson et al., Appl.Phys.Lett. 96, 263511 (2010)
\bibitem{Moore} D.C. Moore, et al., Appl.Phys.Lett. 100, 232601 (2012)
\bibitem{CALDER2015} E.S. Battistelli et al., EPJ C 75, 353 (2015) %doi:10.1140/epjc/s10052-015-3575-6
\bibitem{Doyle} S. Doyle et al., J. Low Temp. Phys. 151, 530 (2008) %doi:10.1007/s10909-007-9685-2
\bibitem{Colantoni} I. Colantoni et al., J. Low Temp. Phys., 184, 131 (2016)  %doi:10.1007/s10909-015-1452-1
\bibitem{Bourrion2011} O. Bourrion et al., JINST 6, P06012 (2011) %doi:10.1088/1748-0221/6/06/P06012
\bibitem{Bourrion2013} O. Bourrion et al., JINST 8, C12006 (2013) %doi:10.1088/1748-0221/8/12/C12006
\bibitem{Casali} N. Casali et al., J. Low Temp. Phys. (2015) %doi:10.1007/s10909-015-1358-y
\bibitem{Cardani2015} L. Cardani et al., Appl. Phys. Lett. 107, 093508 (2015) %doi:10.1063/1.4929977
\bibitem{Cardani2017} L. Cardani et al., Appl. Phys. Lett. 110, 033504 (2017) %arXiv:1606.04565
\bibitem{Ohya} S. Ohya et al., Supercond. Sci. Technol. 27, 015009 (2014) %doi:10.1088/0953-2048/27/1/015009
\bibitem{Catalano} A. Catalano et al., A\&A, 580:A15 (2015)
\bibitem{Vissers} M.R. Vissers et al., Appl. Phys. Lett. 102, 232603 (2013) %doi:10.1063/1.4804286
%\bibitem{CALDERdido} S. Di Domizio et al., J. Low Temp. Phys., 176(3-4):917-923 (2014) %Cryogenic Wide-Area Light Detectors for Neutrino and Dark Matter Searches

\end{thebibliography}
\end{document}